\documentclass[prl,twocolumn,showpacs,floatfix]{revtex4}

\usepackage{amsmath}
\usepackage{amssymb}

\usepackage{graphicx}
\usepackage{dcolumn}
\usepackage{bm}

\begin{document}

\title{Vortex Clusters in Quantum Dots}

\author{H. Saarikoski}
\author{A. Harju}
\author{M.~J. Puska}
\author{R.~M. Nieminen}

\affiliation{Laboratory of Physics, Helsinki University of Technology,
P.O. Box 1100, FIN-02015 HUT, Finland} 

\begin{abstract}

We study electronic structures of two-dimensional quantum dots in
strong magnetic fields using mean-field density-functional theory and
exact diagonalization. Our numerically accurate mean-field solutions
show a reconstruction of the uniform-density electron droplet when the
magnetic field flux quanta enter one by one the dot in stronger
fields. These quanta correspond to repelling vortices forming
polygonal clusters inside the dot.  We find similar structures in the
exact treatment of the problem by constructing a conditional operator
for the analysis.  We discuss important differences and limitations of
the methods used.

\end{abstract}


\date{\today}

\maketitle Vortices appear in many physical systems from tornadoes and
bathtub whirlpools~\cite{tub} to type-II superconductors and rotating
Bose-Einstein condensates~\cite{butts}. In the fractional quantum Hall
effect (FQHE)~\cite{fqhe}, the external magnetic field $B$ penetrates
through the two-dimensional (2D) electron system at the vortex
positions.  Every vortex corresponds to a single magnetic field flux
quantum.  For the quantum Hall state of the filling factor $\nu=1$, a
single vortex is on top of each electron.  For stronger $B$, more
vortices appear and, e.g., the Laughlin state of $\nu=1/3$ attaches 3
vortices on top of each electron.  The vortices keep electrons further
apart, reducing the interaction energy and causing strong correlations
between the electrons.

In this Letter, we report results of detailed numerical investigations
of the electronic structure of 2D quantum dots (QDs)~\cite{reimann} in
strong $B$.  We use both a mean-field and an exact many-body approach.
The $B$ values used are such that our QD states are related to the
FQHE filling $1\le\nu\le 1/3$.  We find both by mean-field and exact
approach that vortices appear one by one inside QD as we strengthen
$B$.  The positions of these vortices are fixed in the mean-field
solutions and visible as zeros in the electron densities.
The vortices form a
polygonal cluster inside the QD.  In the exact treatment the vortices
are mobile. However, by constructing a conditional single-particle
wave function we are able to pinpoint them. Every electron binds one
vortex and there are also additional vortices which are not bound to
any particular electron.  These additional vortices form similar
vortex clusters as found in the mean-field approach. Similarly to the
FQHE, the positions of the additional vortices are such that they
reduce the interaction energy.

We model the 2D QD by an effective-mass Hamiltonian
\begin{equation}
H=\left(\sum^N_{i=1} \frac{(-i\hbar \nabla_i+e {\bf A} )^2}{2 m^*}
+V_{{\rm c}}(r_i)\right) + \frac{e^2}{4\pi \epsilon} \sum_{i<j}
\frac{1}{r_{ij}} \ ,
\label{hamiltonian}
\end{equation}
where $N$ is the number of electrons in the dot, ${\bf A}$ is the
vector potential of the perpendicular $B$, $m^*$ the effective
electron mass, and $\epsilon$ is the dielectric constant of the
medium.  We use a parabolic confinement $V_{{\rm c}}(r) = \frac 12 m^*
\omega_0^2 r^2$, and the material parameters of GaAs, $m^*/m_e=0.067$
and $\epsilon/\epsilon_0= 12.4$. We assume the Zeeman effect to be
strong enough to spin-polarize our system in the $B$ range considered.
We solve the model using both the mean-field density-functional theory (DFT)
and the exact diagonalization (ED).  We use two variants of DFT, namely
the spin-DFT (SDFT) and the current-spin-DFT
(CSDFT)~\cite{vignalerasolt}.  The DFT solutions are found on a
real-space grids without symmetry restrictions~\cite{saarikoski1}.
Special effort is laid on the numerical accuracy and the convergence
is tested with large grids up to the size of $256 \times 256$ grid points.
The exchange-correlation effects are taken into account using local
approximations~\cite{LDA}. In CSDFT, the effect
of currents is also included in the exchange-correlation functionals,
again in a local approximation.  CSDFT is computationally more
demanding than SDFT.  As both schemes give qualitatively similar
results, we have mainly used SDFT.  Our ED calculations use wave
functions restricted to the lowest Landau level
(LLL)~\cite{yang,oaknin,seki96JPSJ}.

The starting point for our study is a $B$ value where the
maximum-density-droplet (MDD) state is formed~\cite{macdonald}.  This
is a finite-size precursor of the $\nu=1$ quantum Hall
state~\cite{fqhe}.  This state can be found in various QD geometries,
and it is characterized by a rather flat and compact electron
density~\cite{Esa}.  In our case, MDD is formed by the LLL orbitals
with angular momentum $l=0,1,\dots,N$, and the total angular momenta
$L$ equals to $N(N-1)/2$. 
In MDD, one vortex is bound to each electron to give the correct fermion
nature in the LLL wave function.

\begin{figure}
\includegraphics[width=.96\columnwidth]{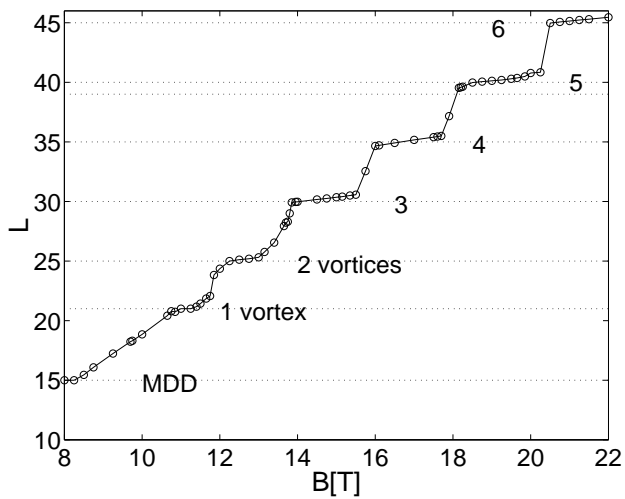}
\caption{Angular momentum $L$ of the six-electron QD from SDFT (open circles).
The plateaus are characterized by vortex-holes in the electron density.
The number next to a plateau gives the number of vortices inside QD.
The horizontal lines correspond to ground state $L$ values from ED.}
\label{fig:lz}
\end{figure}

\begin{figure}
\includegraphics[width=.99\columnwidth]{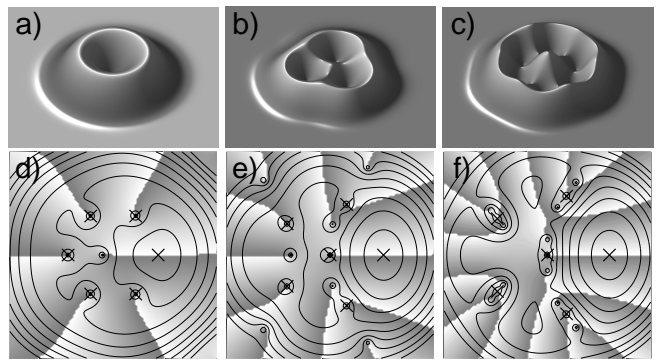}
\caption{Upper panel: Electron densities of the six-electron QD calculated
with CSDFT for single-vortex (a), three-vortex (b), and six-vortex (c)
configurations.  Lower panel: Most probable electron positions
$\{\mathbf{r}_i^*\}_{i=1}^N$ (crosses) and conditional electron
densities (contours) and phases (gray-scale) from ED for the cases of
the upper panel: $L=21$ (d), $L=30$ (e), and $L=45$ (f).  We probe
with the rightmost electron. The densities are on a logarithmic scale
(values $e^{-i}$, $i=1,3,\dots,17$) and the phase changes from $\pi$ to
$-\pi$ on the lines where shadowing changes from the darkest gray to
white.}
\label{fig:vortices}
\end{figure}

Our study focuses on large $B$ values beyond the MDD region, where
ED shows the ground states to occur only at certain ``magic'' $L$ values, 
and $L$ exhibits a
stepwise structure as a function of $B$~\cite{maksym,wojs,seki96JPSJ}.
We show that the ground states in this region can be
characterized by an increasing number of additional vortices
entering the QD and forming vortex clusters.
We first analyze the post-MDD region of a six-electron QD.  Setting the
confinement strength to $\hbar\omega_0=5$~meV, SDFT predicts the MDD
formation at $B\approx 5$~T, and the state is a ground state up to
$B\approx 8$~T~\cite{saarikoski1}.  Fig.~\ref{fig:lz} shows the SDFT
result for $L$ as a function of $B$.  The ED ground-state $L$ values,
which agree with the previous calculations~\cite{wojs}, are given by
the dashed lines.  The values $L=15+6n$, where $n$ is an integer,
reflect the six-fold rotational symmetry whereas values $L=15+5n$
correspond to the five-fold rotational symmetry~\cite{seki96JPSJ}.
The SDFT results show plateaus with a small slope just at the magic
$L$ values.  A possible exception is the $L=39$ magic momentum which
is not clearly visible in the SDFT results.  One should note that the
unrestricted Hartree-Fock approximation has been shown to follow the
trend in angular momentum but not to reproduce the staircase
structure~\cite{peeters}.  As shown by densities in
Figs.~\ref{fig:vortices}(a)--(c), the plateaus are characterized by
vortex-holes, i.e., rotating currents with zero electron density
at the center. The number of vortices increases by
one between plateaus of nearly constant $L$.
There are vortices also further away from the dot center,
where the electron density is a tiny fraction of the maximum density.

To compare our ED results with the DFT ones, we have constructed an
operator that efficiently shows the positions of the vortices in the
system.  We first find the most probable electron positions
$\{\mathbf{r}_i^*\}_{i=1}^N$ by maximizing the density $|\Psi|^2$.
One of the electrons is then moved to a new position $\mathbf{r}$ and
a conditional single-particle wave function is evaluated as
$$
\psi_\mathrm{c}(\mathbf{r})=
\frac{\Psi(\mathbf{r},\mathbf{r}_2^*,\dots,\mathbf{r}_N^*)}
{\Psi(\mathbf{r}_1^*,\mathbf{r}_2^*,\dots,\mathbf{r}_N^*)} \ .
$$ The change in the phase can be obtained from the angle $\theta$ of
the wave function
$\psi_\mathrm{c}(\mathbf{r})=|\psi_\mathrm{c}(\mathbf{r})| \exp(i
\theta(\mathbf{r}))$.  We have plotted $\psi_\mathrm{c}$ in
Figs.~\ref{fig:vortices}~(d)--(f). The electron density is shown using
contours and the phase is shown using a gray-scale to indicate the
angle.  The vortices show up as zeros in $\psi_\mathrm{c}$. A rotation
around one vortex changes the phase by $2\pi$.  One can see that the
number of additional vortices in the inner part of the QD agrees with our DFT
results.  There are also vortices outside the ring of fixed electrons,
as in the CSDFT case.  One should note that in the $L=45$
solution of Fig.~\ref{fig:vortices}~(f), the total number of vortices
close to each fixed electron is three, as in the Laughlin wave
function for the $\nu=1/3$ quantum Hall state~\cite{fqhe}.  Unlike in
the Laughlin state, there is a repulsion between the vortices, forcing
two of them to stay on the opposite sides of the fixed electron. However,
the overlap between the Laughlin and the exact state is high, 0.98.

The electron densities in Figs.~\ref{fig:vortices}(d)--(f) also show
the Wigner molecule formation~\cite{wm}. In Fig.~\ref{fig:vortices}(d)
the conditional density is still well spread over the whole QD (consider,
e.g., the third contour from the top, corresponding to the density
$e^{-5}\approx 0.007$), whereas in Fig.~\ref{fig:vortices}(f), the
density is strongly peaked around the most probable position of the
probe electron. One should note that the most probable electron
positions approach the classical positions in the Wigner-molecule
limit.  For the Coulomb interaction, the classical configuration of six
point charges is a pentagon with one electron in the center.  This
result is in accord also with the CSDFT-solution for the six-vortex
case which shows an electron in the middle and a five-electron ring.
This is also the most probable electron configuration in most ED
ground states, the one in Fig.~\ref{fig:vortices}(d) being one of the
exceptions having the hexagonal configuration. In many ground states,
symmetry eliminates one of the two possible configurations. For the
$L=45$ case of Fig.~\ref{fig:vortices}(f), the hexagonal
configuration has also a high probability, namely of 77~\% of the
pentagonal one, showing that the electrons still make multi-particle
exchanges in this ground state~\cite{wm}.

From the computational point of view, constructing the conditional
wave function $\psi_\mathrm{c}$
is a demanding task. The basic reason for this is that it does {\em
not} contain any integrations of coordinates, which leaves
$\psi_\mathrm{c}$ a true $N$-particle operator. Thus one is forced to
actually construct the Slater determinants in the ED wave function
expansion. We have evaluated the determinants using a LU
factorization.  The computational cost in constructing
$\psi_\mathrm{c}$ for the $L=45$ case with a grid size of
100$\times$100 for $\mathbf{r}$ is 20 times more than that of a mere
diagonalization. For this reason, previous ED works have mainly
concentrated on operators which can be written using one-particle and
two-particle operators~\cite{seki96JPSJ}.

As the Hamiltonian of Eq.~(\ref{hamiltonian}) is rotationally
symmetric, the particle density should also have this property.
Calculations, however, show that the particle density from DFT is {\em
not} necessarily symmetric.  Vortex solutions clearly break the
rotational symmetry of the particle density in the case of more than
one vortex.  Other solutions with broken rotational symmetry are,
e.g., spin-density-wave and charge-density-wave structures
\cite{koskinen97,reimann99}.  The analysis and interpretation of these
solutions are highly non-trivial.  Symmetry-breaking may result from
an unphysical mixture of states that do not mix in an exact
treatment~\cite{hirose,ari}.  However, for the vortex case, there are strong
topological reasons behind the broken symmetry in DFT. Namely, if we
suppose that there are vortices in the true many-body solution of the
system, this directly forces the vortices to localize.  This follows
from considering the Kohn-Sham (KS) equations close to a vortex.  The
kinetic energy diverges if the electron density of the KS orbit does
not vanish at the vortex position.  As a mean-field theory, DFT misses
relative coordinates of the electrons, and thus the vortices must
localize in space. In the true many-body solution of the system,
however, the vortices move as the electron coordinates are
changed. Thus the vortices behave as quantum mechanical particles in
the exact treatment, showing quantum-mechanical zero-point motion, but
the mean-field approximation forces them to behave as classical
particles.

\begin{figure}
\includegraphics[width=.99\columnwidth]{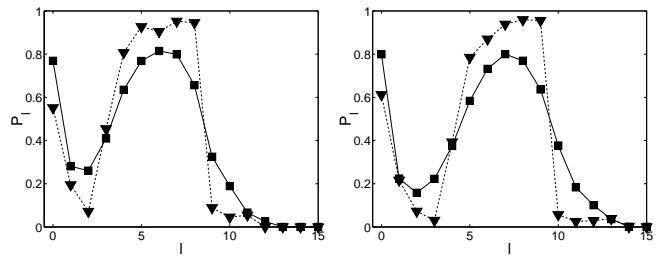}
\caption{Occupation of angular momentum eigenstates for solutions with
three (left) and four vortices (right).  Triangles mark the CSDFT and
squares the ED results, respectively.  The minima at the low $l$ values show the formation of vortices in the QD.  }
\label{fig:projections}
\end{figure}

The CSDFT and ED solutions are analyzed further by calculating the
occupation of the angular-momentum eigenstates, i.e., the projections
on the Fock-Darwin orbitals. For the MDD, all angular momenta $l\le
N-1$ have occupancy one and the others are zero. The single-vortex
state of DFT has the $l=0$ orbital unoccupied and the following $l=1,
\dots, 6$ orbitals occupied.  Fig.~\ref{fig:projections} shows the
occupations for three- and four-vortex solutions ($L=30$ and $L=35$,
respectively).
The missing occupation at low $l$ increases with the number of
vortices. Thus vortices appear as holes in the MDD.  As the magnetic
field squeezes the QD, it becomes at some points favorable to extend
the system by creating an additional vortex-hole into the
system~\cite{yang}.  From the occupation in Fig.~\ref{fig:projections}
one can see that the DFT vortices are more localized to certain
angular momentum value than the ED ones. The SDFT occupations differ
more from the ED results than the CSDFT ones.

We have performed SDFT calculations also for a 24-electron QD, with
the confinement strength changed to $\hbar \omega_0= 1.8940$~meV.  The
calculations predict the formation of clusters of vortices in high
magnetic fields. Fig.~\ref{fig:n24} shows the electron density at 5 T
($L\approx 491$).  The currents in our DFT solutions
circulate counterclockwise on the edge of the QD whereas the
circulation is {\em clockwise} around the vortices.  This behavior is
consistent with the classical picture of a conducting ring where the
inner circulation of electrons reverses the current near the center
hole~\cite{lent}.  Since the vortex holes behave as classical
repelling particles in DFT, the vortex clusters inside the
QD are usually similar to the configurations of classical
point charges confined by a parabolic external potential~\cite{bedanov}.
For instance, the solution at 5 T contains 14 vortices arranged in the
(4,10) configuration, i.e., four vortices in the middle and ten in the
second shell, which is the same as the ground state configuration for
14 classical particles.  The positions of vortices are not independent
of the electron degrees of freedom and therefore the vortex
configurations do not necessarily match those of the classical case.
This can be seen in Fig. {\ref{fig:vortices}(c) where the vortex
cluster is hexagonal in contrast to the (1,5) configuration for the
classical 6-particle system.
\begin{figure}
\includegraphics[width=.96\columnwidth]{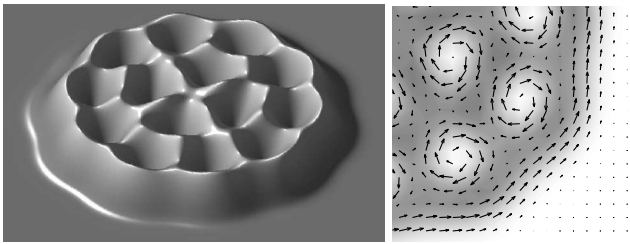}
\caption{Left panel: SDFT electron density of a 24-electron QD at 5~T
showing 14 vortices. Right panel: Electron density (gray-scale,
the white regions have low electron density and correspond to vortices)
and currents (arrows) on the edge of the QD.}
\label{fig:n24}
\end{figure}

In Bose-Einstein condensates, there is a repulsive interaction between
the atoms and the many-body wave function for a rotating condensate
can be constructed similarly to that for a 2D electron gas in a magnetic
field~\cite{manninen01}.  This means that the two systems are
analogous and explains why the mean field solution for the rotating
Bose-Einstein system also shows vortex clusters \cite{butts}.

The DFT solutions between the plateaus show charge-density-wave (CDW)
solutions with a fractional $L$. In the first CDW domain $L$ increases
linearly with $B$ from $15$ to $21$ (see Fig. \ref{fig:lz}).  The
particle density in this region shows six charge maxima in the form of
a hexagon and with currents flowing {\em counterclockwise} around
them. The radius of each charge lump is of the order of the magnetic
interaction length $\ell_B=\sqrt{\hbar/eB}$. These solutions are
reminiscent of the localized electron states found by Reimann {\em et
al.} just above the MDD region~ \cite{reimann99}.  Reimann {\em et al.} do
not show results for the higher $B$ values where vortex solutions
appear in our calculations.  We want to underline that finding the
vortex solutions requires high numerical accuracy.  Our real-space
scheme is superior to the plane-wave expansion of wave functions used
by Reimann {\em et al.}, especially in describing the vanishing
electron density at the vortex core.

We have analyzed the angular momentum occupations of the first CDW
region between $15$ and $21$.  The results show that these states are
combinations of the $L=15$ and $L=21$ states.  A mixture of these two
states results in a CDW with six peaks in form of a hexagon.  This
result is in disaccord with the (excited state) ED solutions between
$L=15$ and $21$ which show a vortex-hole moving from the outer edge
toward the center~\cite{oaknin}. The DFT mixing of eigenstates
belonging to different angular momentum eigenvalues can be thought to
be unphysical and resulting from limitations of DFT. These
limitations can be seen also in the mixing of total spin states in
SDFT~\cite{ari}.  Our analysis of the symmetry-breaking solutions
shows that DFT, which is an indispensable tool for large systems,
may reveal proper correlations between the electrons via these
solutions.  But the correct interpretation of the results requires
careful analysis completed by, e.g., the ED method.

To conclude, we have found vortex clusters as mean-field solutions of
the two-dimensional quantum dots in strong magnetic fields. The
external magnetic field penetrates through the dot at the position of
the vortices, inducing electron currents that circulate around
them. The exact treatment of the problem shows similar features in the
conditional electron wave function. The mean-field approach has shown
to lack the quantum-mechanical nature of the vortices, leading to a
vortex localization and a broken rotational symmetry for the cases
with more than one additional vortex. Apart from these facts, DFT 
results have shown to accurately describe the electronic
structure of this challenging system, enabling one to study system
sizes beyond reach of the more exact treatments.  We hope that this
theoretical prediction would inspire further work for actual
experimental evidence of vortex formation in quantum dots.  This could
presumably be seen indirectly in the magnetization measurements of
quantum dots.  A more direct observation could be made if the vortices
get localized due to a lower symmetry of the system, e.g., in quantum
dots distorted by impurities.  The stability and structure of vortices
in non-symmetric potentials is, however, a nontrivial problem and
further theoretical work is needed.

\acknowledgments

We thank S.~M. Reimann and E. R\"as\"anen for fruitful
discussions.  This work has been supported by Academy of Finland
through the Centre of Excellence Program (2000-2005).

\end{document}